\documentclass[11pt]{article}
\usepackage{moriond,epsfig}

\def\be{\begin{equation}}
\def\ee{\end{equation}}
\def\bea{\begin{eqnarray}}
\def\eea{\end{eqnarray}}

\begin{document}
\vspace*{3cm}
\title{BRANON DARK MATTER}

\author{ J.A.R. CEMBRANOS$^{1,2}$, A. DOBADO$^2$, A.L. MAROTO$^2$ }

\address{$^1$ Departamento de Estad\'{\i}stica e
Investigaci\'on Operativa III,\\
$^2$ Departamento de  F\'{\i}sica Te\'orica,\\
 Universidad Complutense de
  Madrid, 28040 Madrid, Spain}
\maketitle\abstracts{
In the brane-world scenario, our universe is understood as
a three dimensional hypersurface  embedded  in a higher
dimensional space-time. The fluctuations of the brane along
the extra dimensions are seen from the four-dimensional point
of view as new fields whose properties are determined by the
geometry of the extra space. We show that such
branon fields can be massive, stable and weakly interacting, and
accordinlgy they are natural  candidates to explain the 
universe missing mass problem. We also 
consider the possibility of producing branons non-thermally
and their relevance in the cosmic coincidence problem.
Finally we show that some 
of the branon distinctive signals could be detected
in future colliders and in direct or indirect dark matter searches.  }

\section{Introduction}
The construction of extra-dimensional models has been
revived in recent years within  the so called brane-world scenario \cite{ADD}.
The main assumption of this scenario is that 
by some (unknown) mechanism, matter fields are constrained to live
in a three-dimensional hypersurface (brane) embedded in the higher
dimensional (bulk) space. Only gravity is able to propagate in 
the bulk space, but the fundamental scale of gravity in $D$ 
dimensions
$M_D$ can be much lower than the Planck scale, the volume
of the extra dimensions being responsible for the actual value
of the Newton constant in four dimensions.
The fact that rigid objects are incompatible with General 
Relativity
 implies
that the brane-world must be  dynamical and can move and
fluctuate along the extra dimensions. Branons are
precisely the fields parametrizing the
position of the brane in the extra coordinates \cite{GB,DoMa}.
Thus, in four dimensions 
branons could be
detected through their contribution to the 
induced space-time metric. 

\section{Branon dark matter}

Let us consider  our four-dimensional space-time 
$M_4$ to be
embedded in a $D$-dimensional bulk space whose coordinates 
will be denoted by $(x^{\mu},y^m)$, where 
$x^\mu$, with $\mu=0,1,2,3$, correspond to the 
ordinary four dimensional space-time and $y^m$, with 
$m=4,5,\dots,D-1$, are coordinates of the compact extra
space of typical size $R_B$.
For simplicity we will assume that the bulk 
metric tensor takes the following form:
\begin{eqnarray}
ds^2=\tilde g_{\mu\nu}(x)W(y)dx^\mu dx^\nu- g'_{mn}(y)dy^m dy^n
\label{metric}
\end{eqnarray}
where the warp factor is normalized as $W(0)=1$. 
The position of the brane in the bulk can be parametrized as
$Y^M=(x^\mu, Y^m(x))$, and  we assume for simplicity that the ground
state of the brane corresponds to $Y^m(x)=0$. 

In the simplest case in which the metric is not
warped along the extra dimensions, 
i.e. $W(y)=1$,  
the transverse brane fluctuations are massless and
they can be parametrized by the Goldstone boson fields 
$\pi^\alpha(x),\; \alpha=4,5, \dots D-1$, 
associated to the spontaneous breaking of the extra-space 
traslational symmetry.  
 In that case we can choose the $y$
coordinates   so that the branon fields are
proportional to the extra-space coordinates:
$\pi^\alpha(x)
=f^2\delta_m^\alpha Y^m(x)$, 
where the proportionality constant is related to the brane 
tension $\tau=f^4$.

In the general case, the curvature generated by the
warp factor explicitly breaks the traslational 
invariance in the extra space. Therefore branons  acquire a mass 
matrix which is given precisely by the bulk Riemann tensor
evaluated at the brane position: 
$M^2_{\alpha\beta}=\tilde g^{\mu\nu}R_{\mu\alpha\nu\beta}\vert_{y=0}$.

The dynamics of branons can be obtained from the 
 Nambu-Goto action.
In addition, it is also possible to get their couplings to
the ordinary particles just by replacing the space-time by
the induced metric in the Standard Model (SM) action.
Thus we get up to quadratic terms in the branon fields \cite{GB,DoMa,BSky}: 
\begin{eqnarray}
S_{Br} 
&=&\int_{M_4}d^4x\sqrt{\tilde g}\left[\frac{1}{2}
\left(\tilde g^{\mu\nu}\partial_{\mu}\pi^\alpha
\partial_{\nu}\pi^\alpha
-M^2_{\alpha\beta}\pi^\alpha \pi^\beta
\right)
+\frac{1}{8f^4}\left(4\partial_{\mu}\pi^\alpha
\partial_{\nu}\pi^\alpha-M^2_{\alpha\beta}\pi^\alpha \pi^\beta 
\tilde g_{\mu\nu}\right)
T^{\mu\nu}_{SM}\right]\nonumber \\
\label{Nambu}
\end{eqnarray}
\vspace*{-1.0cm}
\begin{figure}[h]
\centerline{\epsfxsize=10cm\epsfbox{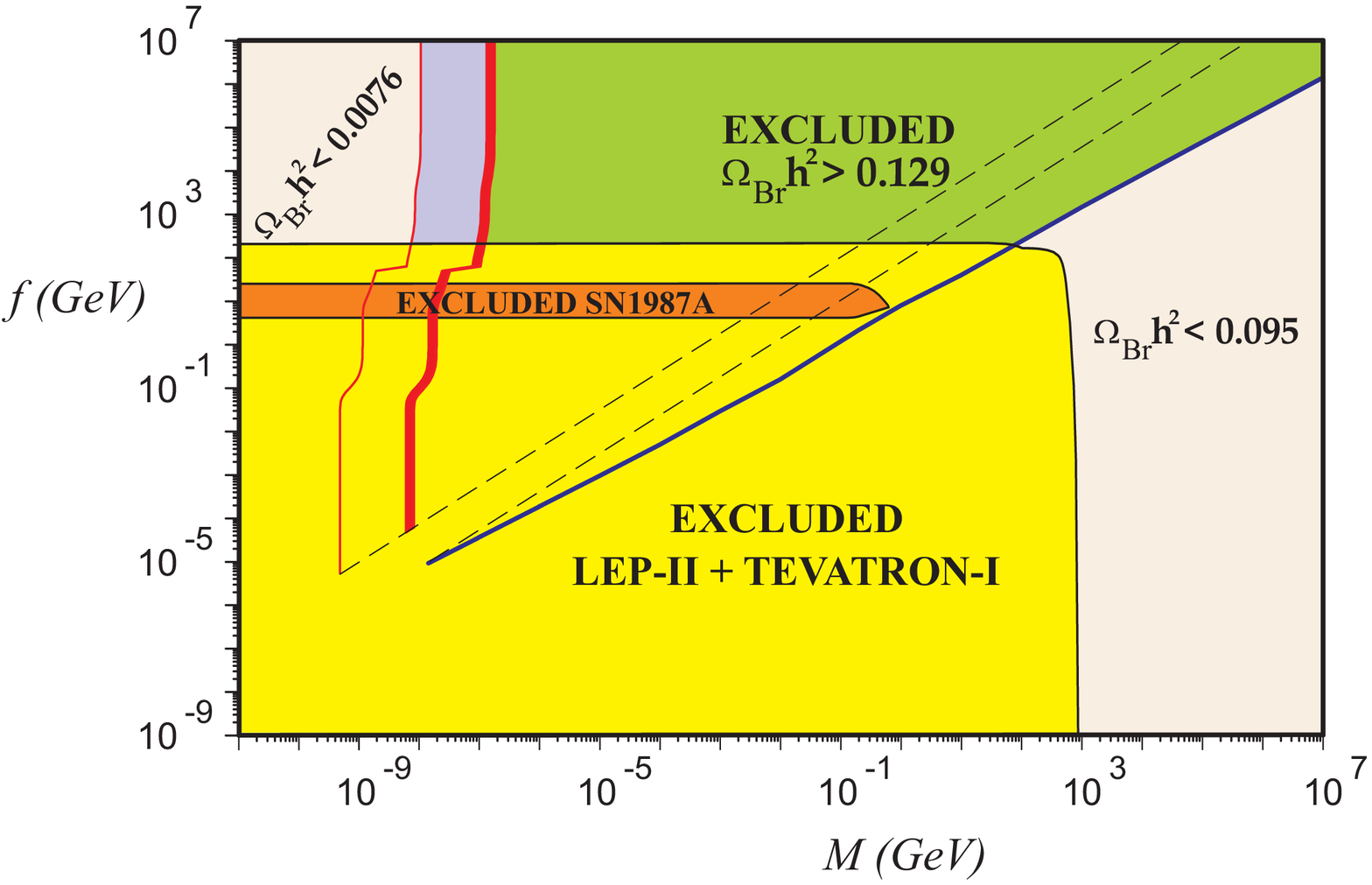}}
\end{figure}

\noindent
{\footnotesize Figure 1: Relic abundance in the $f-M$ plane for a 
model with one branon of
mass: $M$. The two lines on the left correspond to the 
$\Omega_{Br}h^2=0.0076$ and 
$\Omega_{Br}h^2=0.129 - 0.095$ curves for hot-warm relics, 
whereas the right line
corresponds to the latter limits for cold relics (see \cite{CDM} 
for details). 
The lower area is excluded by single-photon processes at LEP-II 
\cite{ACDM} together 
with monojet signal at Tevatron-I \cite{ACDM}. The astrophysical 
constraints are less
restrictive and they mainly come from supernova cooling by branon 
emission 
\cite{CDM}.}
\vspace*{0.2cm}

We can see that branons  interact with the  SM
 particles through their energy-momentum tensor.
The couplings are controlled by the brane 
tension scale $f$. For large $f$, branons are therefore 
weakly interacting particles. The sign of the branon fields is determined by the orientation
of the brane submanifold in the bulk space. Under a parity 
transformation on the brane ($x^i\rightarrow -x^i$), the orientation
of the brane changes sign provided the ordinary space has an odd number
of dimensions, whereas it remains unchanged for even spatial dimensions.
In the case in which we are interested with three ordinary spatial
dimensions, branons are therefore pseudoscalar particles.
Parity on the brane then requires that branons always couple 
to SM particles by pairs, which ensures that they are 
stable particles. This fact means that branons are natural dark matter candidates.

When the branon annihilation rate, $\Gamma=n_{eq}\langle\sigma_A v
\rangle$,                                 
equals the universe expansion rate $H$, the branon abundance 
freezes out relative to the entropy density. This happens at the 
so called freeze-out temperature $T_f$. We have computed this 
relic branon abundance in two cases: either relativistic branons at 
freeze-out (hot-warm) or non-relativistic (cold), and assuming that 
the evolution of the universe is standard for $T<f$ (see Fig. 1).

\section{Non-thermal branon production and the cosmic coincidence
problem}

So far we have considered only the thermal production of branons.
However, if the maximum temperature reached in the universe is
smaller than the branon freeze-out temperature, but larger than
the explicit symmetry breaking scale, then branons can be 
considered as massless particles decoupled from the rest of 
matter and radiation. In such a case there is no reason to expect
that initially the brane was located at the potential minimum $Y_0=0$,
but in general we will have $Y_0\simeq R_B$, i.e. $\pi_0\simeq f^2R_B$.
As the universe expands, the brane can
start  oscillating around the minimum, and the energy
density of the oscillation  can be seen as
cold dark matter from the brane point of view. This is completely
analogous to the misalignment mechanism for axion production. 
In the case in which $H(T)>\Gamma(T)$, with $\Gamma(T)$ the total
branon annihilation rate, the amplitude of the oscillations is
only damped by the Hubble expansion, but not by particle production.
The corresponding present energy density  would be given by \cite{NT}:
\begin{eqnarray}
\Omega_{Br}h^2\simeq  \frac{6.5\cdot 10^{-20}N}{\mbox{GeV}^{5/2}}
f^4\,R_B^2\,M^{1/2},
\label{nonthermal}
\end{eqnarray}
where $N$ is the number of branon species. 
It is interesting to estimate  typical values for $\Omega_{Br}h^2$
generated by this mechanism. Thus consider the simplest 
non-trivial model in six dimensions
in which the bulk space only contains a 
(negative) cosmological constant $\Lambda_6$ (AdS$_6$ soliton). 
The solutions of Einstein equations in  the
 case in which the extra space has azimuthal symmetry and 
the metric depends only
on the radial coordinate $\rho$ with a periodic angular coordinate
$\theta\in [0,2\pi)$, is given by:
\begin{eqnarray}
ds^2=M^2(\rho)\eta_{\mu\nu}dx^\mu dx^\nu-d\rho^2-L^2(\rho)d\theta^2,
\end{eqnarray}
where, with $k=\sqrt{-5\Lambda_6/(8M_6^4)}$:
\begin{eqnarray}
M(\rho)=\cosh^{2/5}(k\rho);\;\;\;
L(\rho)=\frac{\sinh(k\rho)}{k\, \cosh^{3/5}(k\rho)}. 
\end{eqnarray}

Notice that we have assumed that the presence of the brane
has no effect on the bulk metric. However, even if we include the 
jump conditions at the brane position, it can be seen that the only
consequence would be the introduction of a deficit angle in
the $\theta$ coordinate, which is related
to the brane tension. In addition, in order to compactify the
extra dimensions, it has been shown  that it is 
possible to truncate the extra space by introducing  a 4-brane
at a finite distance $\rho=R_B$ with an anisotropic energy-momentum tensor.
The corresponding branon mass is given by 
$M^2=8k^2/5=-\Lambda_6/M_6^4$.
 
 Let us assume that there is only
one fundamental scale in the theory which is close to 
the electroweak scale,
i.e. we will have $f\sim M_6\sim 1$ TeV, and also assume that
the order of magnitude of the bulk cosmological constant is fixed
by bulk loop effects i.e. $\Lambda_6\sim R_B^{-6}$. In this case, the
branon mass is $M\sim 10^{-33}$ eV, and, in order to recover
the usual four dimensional Planck scale, the size of the extra 
dimension should be $R_B^{-1}\sim 10^{-3}$ eV. Substituting these
values into Eq.(\ref{nonthermal}), we
get $\Omega_{Br}h^2\simeq$ 0.1, in agreement with observations.
Six dimensional models like the one above have been studied also 
in the context
of the dark energy problem. It has been shown that integrating the
volume of the extra space, the natural
value for the brane cosmological constant would be $\Lambda_4\sim 
R_B^{-4}\sim (10^{-3} \mbox{eV})^4$, also in agreement with 
observations. Since the own brane tension does not
contribute to the brane cosmological constant in the $D=6$ case, 
it has been suggested that the amount of fine tuning needed to solve
the dark energy problem would be reduced in these models. 
In addition, as shown above the correct value of the dark matter
energy density can also be obtained without including additional
mass scales in the theory. Therefore,  we see that 
in 6D brane
world models the two fine-tuning problems, i.e. 
the gauge hierarchy and the cosmic coincidence can be related
to a single one, namely, the existence of large extra dimensions 
\cite{NT}.   

\section{Branon searches}

If branons make up the galactic halo, they could be detected by direct search 
experiments from the energy transfer in elastic collisions with nuclei of a 
suitable target. For the allowed parameter region in Fig. 1, branons cannot
be detected by present experiments such as DAMA, ZEPLIN 1 or EDELWEISS.
However, they could be observed by future detectors such as CRESST II, CDMS or
GENIUS \cite{CDM}.

Branons could also be detected indirectly: their annihilations in the galactic 
halo can give rise to pairs of photons or $e^+ e^-$ which could be detected by  
$\gamma$-ray telescopes such as MAGIC or GLAST or antimatter detectors 
(see \cite{CDM} for an estimation of positron and photon fluxes from
branon annihilation in AMS). 
Annihilation of branons trapped in the center of the sun or the earth can 
give rise to high-energy neutrinos which could be detectable by high-energy 
neutrino telescopes such as AMANDA, IceCube or ANTARES.
These searches complement those in high-energy particle colliders 
(both
in $e^+ e^-$ and hadron colliders) in which 
real (see Fig. 1) and virtual branon effects could be measured 
\cite{CDM}.
\vspace*{0.1cm}

{\bf Acknowledgments:} This work is supported by DGICYT (Spain) under project 
numbers FPA 2000-0956 and BFM 2002-01003

\end{document}